\documentclass{webofc}
\usepackage[varg]{txfonts}   % Web of Conferences font
\usepackage{appendix}
\usepackage{amsmath}
\usepackage{multirow}
\usepackage{hyperref}
\usepackage{natbib}
\usepackage{listings}
\usepackage{xcolor}

\definecolor{delim}{RGB}{20,105,176}
\definecolor{numb}{RGB}{106, 109, 32}
\definecolor{string}{rgb}{0.64,0.08,0.08}

\lstdefinelanguage{json}{
    numbers=left,
    numberstyle=\small,
    frame=single,
    rulecolor=\color{black},
    showspaces=false,
    showtabs=false,
    breaklines=true,
    postbreak=\raisebox{0ex}[0ex][0ex]{\ensuremath{\color{gray}\hookrightarrow\space}},
    breakatwhitespace=true,
    basicstyle=\ttfamily\small,
    upquote=true,
    morestring=[b]",
    stringstyle=\color{string},
    literate=
     *{0}{{{\color{numb}0}}}{1}
      {1}{{{\color{numb}1}}}{1}
      {2}{{{\color{numb}2}}}{1}
      {3}{{{\color{numb}3}}}{1}
      {4}{{{\color{numb}4}}}{1}
      {5}{{{\color{numb}5}}}{1}
      {6}{{{\color{numb}6}}}{1}
      {7}{{{\color{numb}7}}}{1}
      {8}{{{\color{numb}8}}}{1}
      {9}{{{\color{numb}9}}}{1}
      {\{}{{{\color{delim}{\{}}}}{1}
      {\}}{{{\color{delim}{\}}}}}{1}
      {[}{{{\color{delim}{[}}}}{1}
      {]}{{{\color{delim}{]}}}}{1},
}
\begin{document}

\title{Data and Decision Traceability for SDA TAP Lab's Prototype Battle Management System}

\author{Latha Pratti, Samya Bagchi, Yasir Latif - Space Protocol LLC,\\
%110 16th Street Suite 1460, Denver CO 80202\\
\texttt{\{latha,samb,yasir\}@spaceprotocol.org}}    % etc.

\abstract{%
  
}
\maketitle
\begin{figure}[h]
    \centering
    \fbox{
    \includegraphics[width=0.9\linewidth]{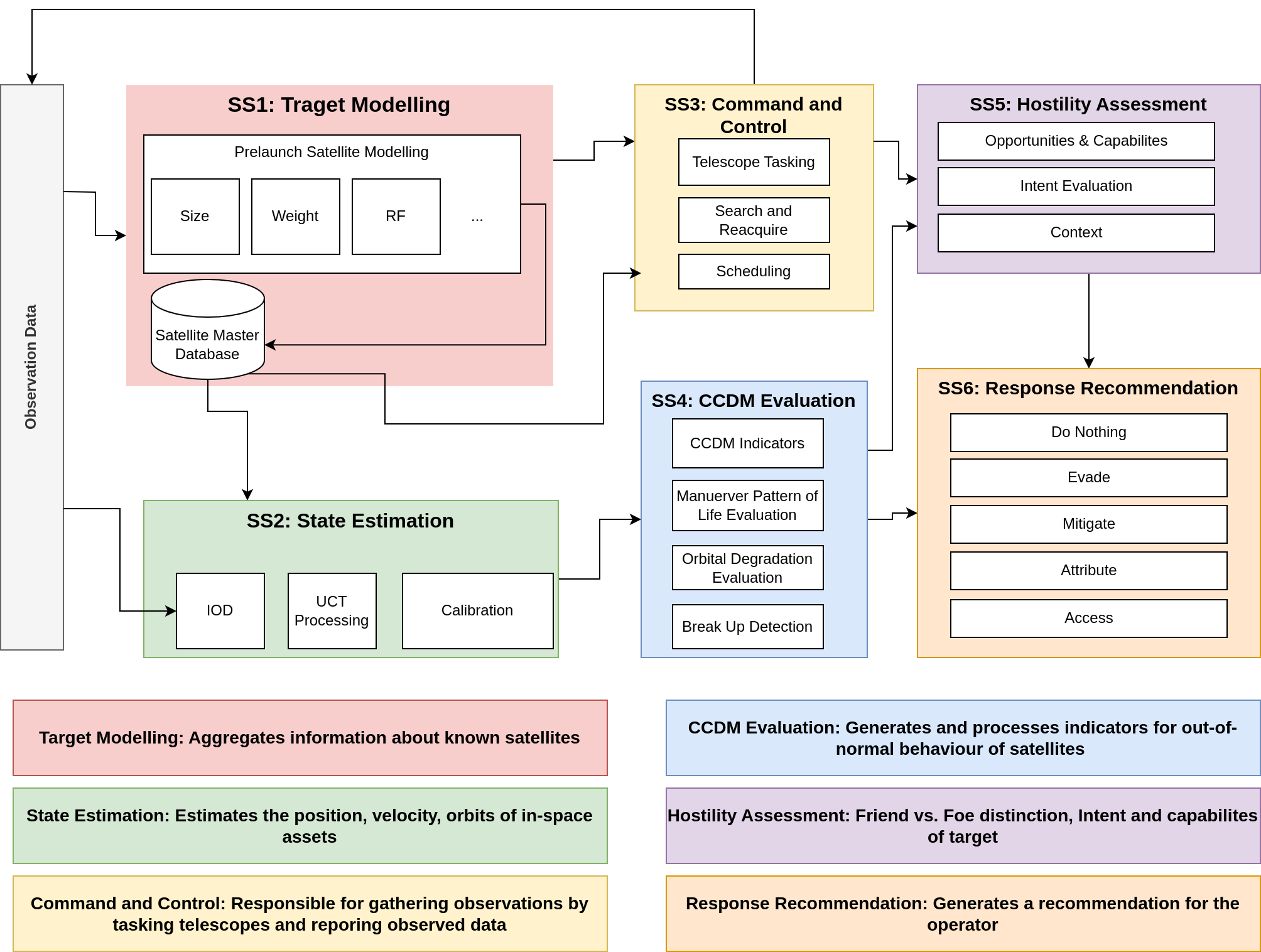}
    }
    \caption{The Prototype Battle Management System (BMS) grouped into its functional subsystems. }
    \label{fig:WA}
\end{figure}
\section{Introduction}
\label{sec:introduction}
 A complex battle management system (BMS) for operator recommendation is being developed for the US Space Systems Command (USSSC) under the Space Domain Awareness (SDA) Tool and Processing (TAP) Lab's Apollo Accelerator program. The BMS ingests sensor input, historic data and related SDA information, conducts hostility assessment of Resident Space Objects (RSOs), and provides response recommendations for operator use. The Apollo accelarator is currently in its sixth cohort (Feb - April 2025) and consist of more than 80 participants from industry, government and academia collaboratively providing algorithms for the BMS system.
 
 The current design of the BMS system (see Figure.~\ref{fig:WA}) consists of seven subsystems with complex data interactions between them, however, it provides opaque recommendations: the operator only sees the response recommendation and is unaware of the complex sequence of decisions made at each step by a different algorithm that lead to the recommendation. This might be acceptable behavior if the system is guaranteed to perform optimally; however, in cases where an intermediate error in the pipeline leads to an incorrect recommendation, detailed auditing of the system, replay of the sequence of events, and fine-grained introspection are required to isolate the source of error within this complex system.

\begin{figure}
    \centering
    \includegraphics[width=0.8\linewidth]{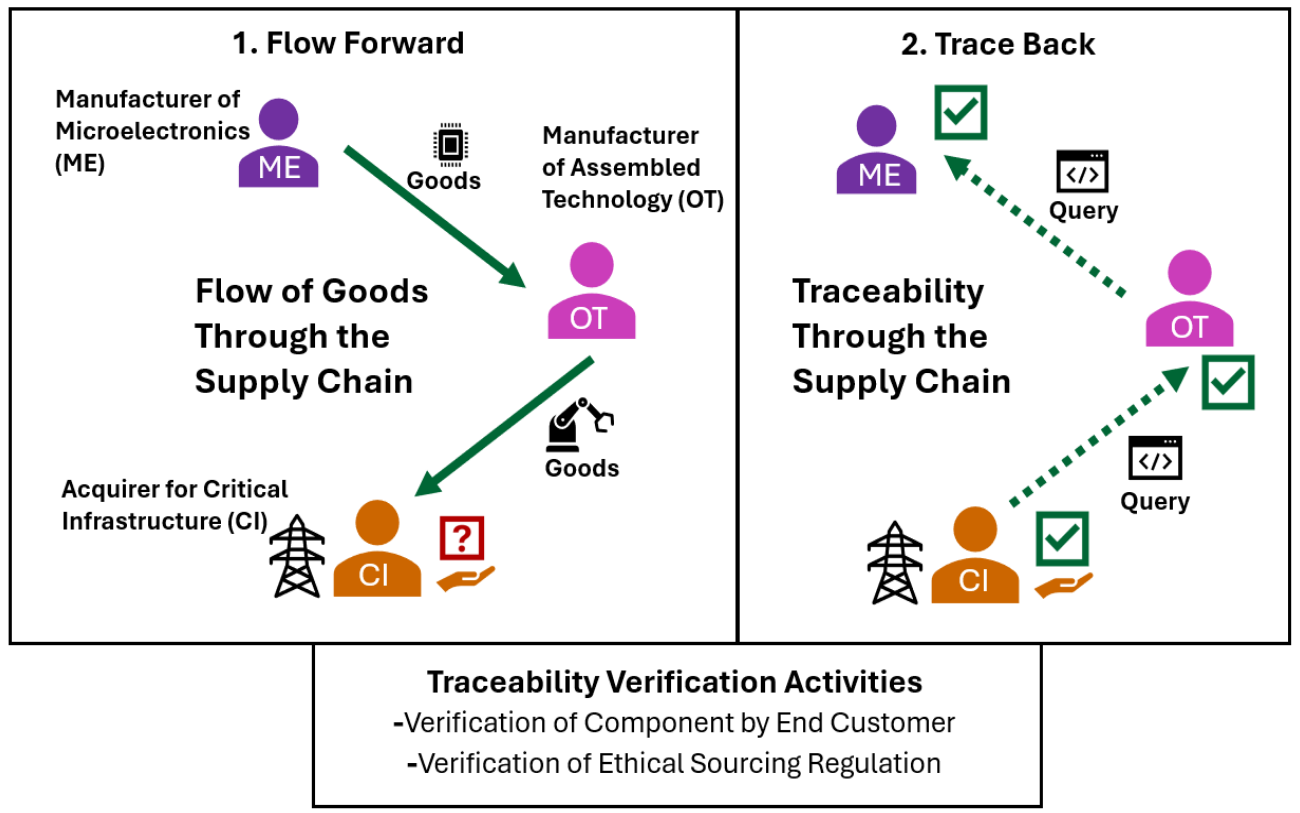}
    \caption{Flow of goods (forward through time) and traceability (backwards through time): Use case of information verification within a manufacturing supply chain. (Figure from~\cite{Pease2024})}
    \label{fig:supply_chain}
\end{figure}

% \section{Supply Chain Traceability and the Target Model Database (TMDB)}
% \label{sec:traceability}

% Within the realm of the SDA TAB Lab, this framework is directly applicable to the Target Model Database (TMDB), which encompasses comprehensive details regarding space assets. Our initial proposal was to investigate the feasibility of establishing a manufacturing supply chain for satellites that would integrate with the TMDB. Nevertheless, the acquisition of such data from public sources quickly emerged as an insurmountable obstacle in this endeavor and, we subsequently explored applying the same core principles of the NIST framework to data and decision traceability within the Welder's Arc system.
%For bibliography use \cite{RefJ}

\section{Data and Decision Traceability}\label{sec:traceability}

Data and decision traceability refers to the comprehensive and auditable documentation of the lifecycle of data, encompassing its origin, transformation, movement, and utilization, as well as the rationale and influencing factors behind decisions using that data.  It represents a systematic approach to recording and linking data elements and decision points, creating a verifiable chain of evidence.  This chain allows for the reconstruction of data provenance and the justification of choices made throughout a system's operation.  

In complex systems, particularly those involving multiple stakeholders or automated decision-making processes, traceability becomes crucial for establishing accountability, ensuring data integrity, facilitating debugging and optimization, and supporting compliance with regulatory requirements.  Effective traceability mechanisms enable the identification of data lineage, the assessment of data quality, and the analysis of decision-making processes, ultimately contributing to increased transparency, trust, and understanding.  Furthermore, it provides a foundation for continuous improvement by enabling the identification of potential biases, inefficiencies, and areas for refinement in both data handling and decision-making algorithms.

\begin{figure}[h]
    \centering
    \includegraphics[width=\linewidth]{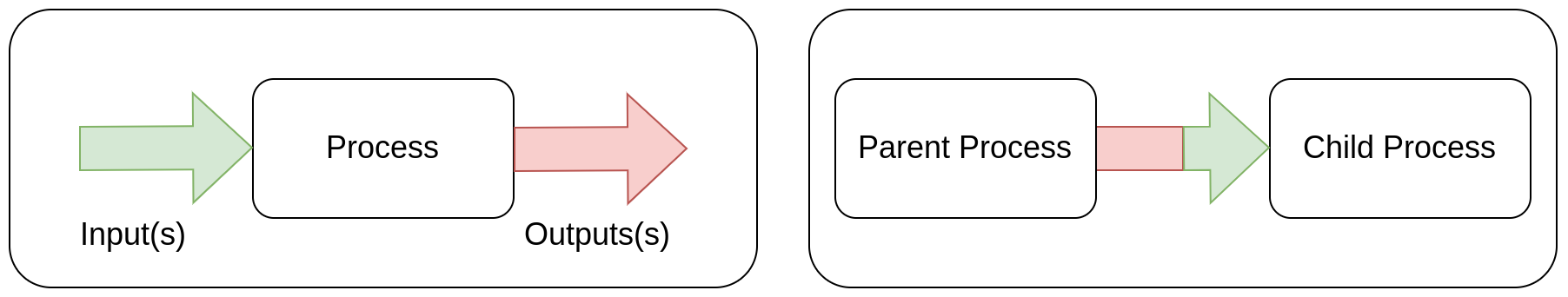}
    \caption{Traceability aims to capture \textbf{Left:} For each process (algorithm) in the system, the instantaneous inputs and outputs and \textbf{Right:} Establishing causal relationship between processes by tracking when one (child) consumes the output of another (parent). These two simple rules allow modeling complex systems with hundreds of algorithms and complicated data exchange patterns.}
    \label{fig:what_is_traceability}
\end{figure}

Information needed to provide data and decision traceability consists of two main components as represented in Fig.~\ref{fig:what_is_traceability}. For each process in the system, the information captured (inputs) and generated (outputs) needs to tracked. To provide traceability of how information flows through the system, algorithms are connected based on the output-input connections.  Further details are presented in the following section with an example shown in Sec.~\ref{sec:traceability_example}.

\section{Decision Traceability for the BMS system}\label{sec:traceability_bms}

Space Protocol has been actively involved with MITRE Corporation since May 2024 and has been collaborating on the implementation of their NIST IR 8536 ``Supply Chain Traceability: Manufacturing Meta-Framework''~\cite{Pease2024}.  The comprehensive framework is designed to enhance traceability across manufacturing supply chains, focusing on improving product provenance, pedigree, and supply chain transparency. The Meta-Framework organizes, links, and enables querying of traceability data across manufacturing supply chains (Figure.~\ref{fig:supply_chain}).

Space Protocol is applying the principles derived from \cite{Pease2024} to the BMS system to achieve introspection, auditing, and replay of data and decisions that ultimately lead to a response recommendation. The core goal of \textbf{decision traceability} is to ensure transparency, accountability, and integrity within the BMS system. This is accomplished by providing a clear, auditable path from the system's inputs all the way to the final decision. This traceability enables the system to track the various algorithms and data flows that have influenced a particular outcome. 

The incorporation of the message bus as the common communication framework, as opposed to the earlier model of end-to-end communication between different algorithms via API calls, has opened up the possibility of end-to-end data and decisions traceability within the system. Instead of direct API calls, different algorithms now send messages to the message bus to make their contributions known. Each message consists of a well-defined schema, that captures key information about data, its origin and the process that generates it (see Listing.~\ref{lst:header} for an example). Using the message bus as the common communication framework enables every component of the system to contribute details about its inputs, outputs, and metadata with each interaction. This data is collected as the system evolves, providing a chronological log of all the data transformations and decisions over time.

\begin{figure}[h]
\centering
\includegraphics[width=\textwidth]{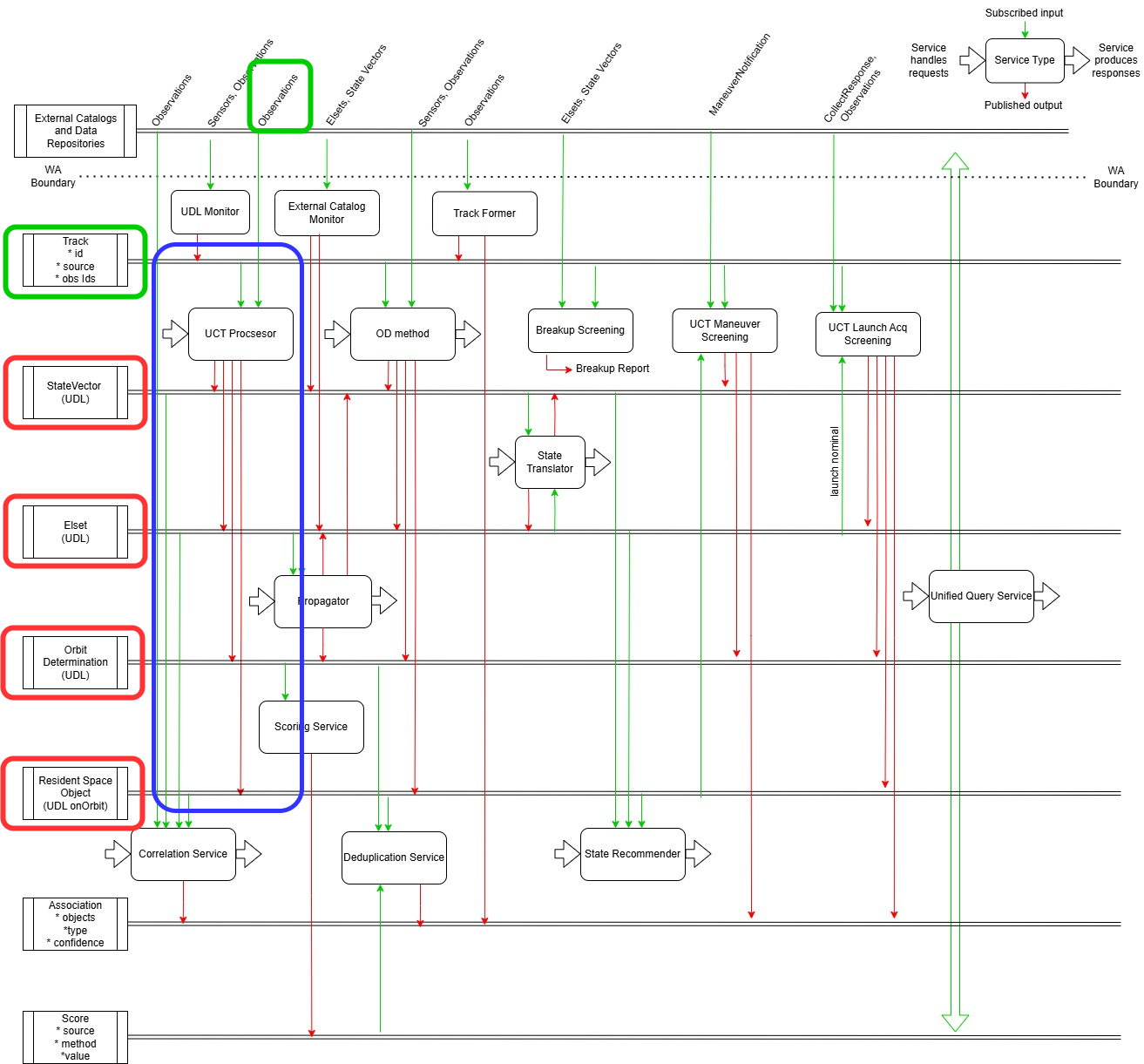}
\caption{Data required for traceability: Inputs (green) and outputs (red) of an algorithm are captured via the header schema along with metadata  for a UCT processing algorithm (blue box).}
\label{fig:UCT}
\end{figure}

The message bus serves as the central mechanism for collecting the data needed for traceability. It is being developed through a collaboration between Millennial Software and Pacific Northwest National Labs (PNNL), with Space Protocol contributing to the design of the message headers that enable the traceability. This collaboration ensures the framework supports all components within the BMS system.

An example required for traceability is shown in Figure~\ref{fig:UCT} where a UCT processing algorithm consumes \texttt{tracks} and \texttt{observations} and generates four distinct outputs. The underlying message bus schema enables collection of metadata about each unique input and output of the algorithms. When needed, any decision (output) can be traced back to its immediate parents, which in turn are related to their parents, building up the complete traceability chain that provides data and decision visibility. 

\subsection{Traceability: Object Discovery}\label{sec:traceability_example}

\begin{figure}
    \centering
    \includegraphics[width=0.8\linewidth]{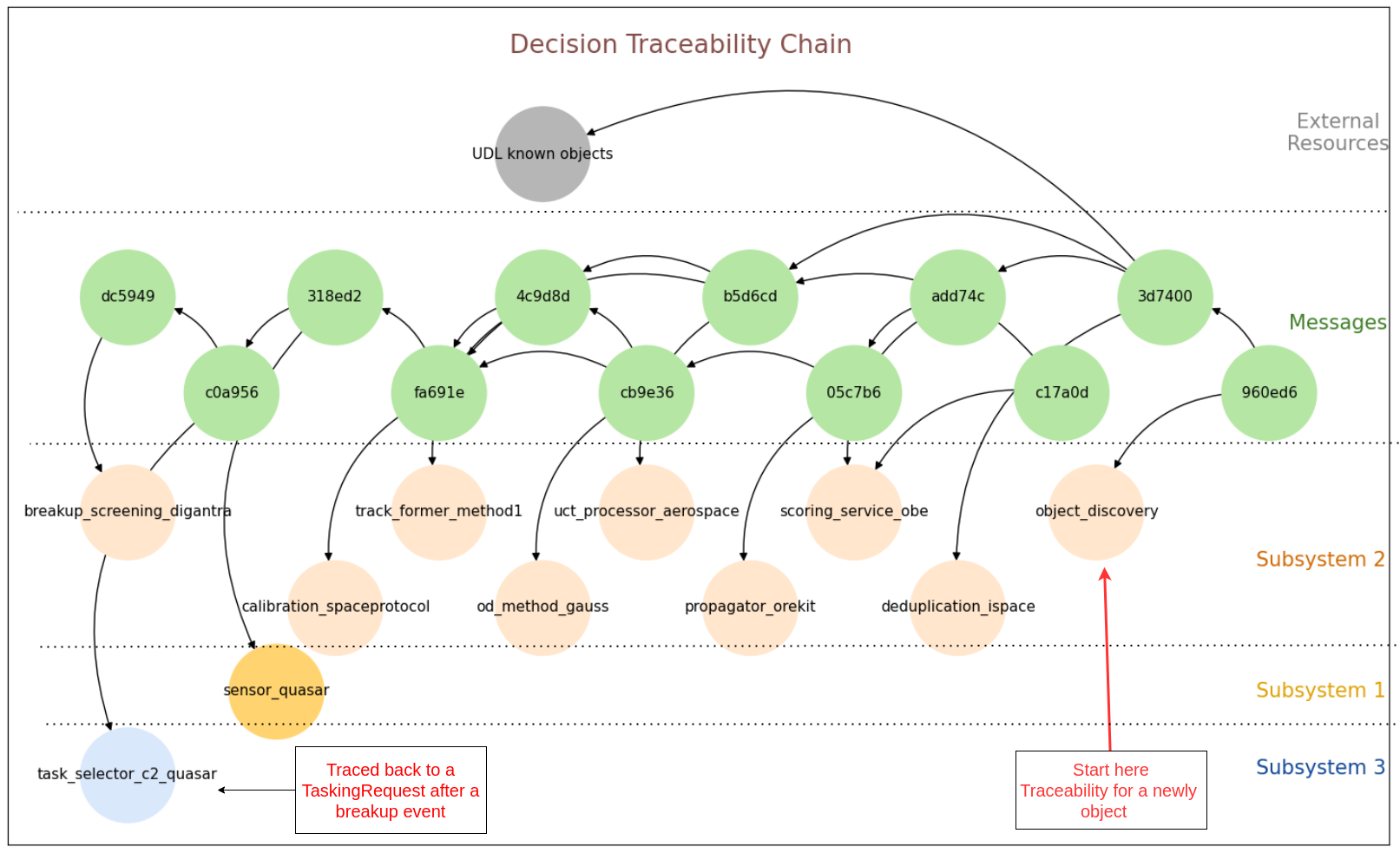}
    \caption{Traceability links a newly discovered object to the breakup event in which it originated.}
    \label{fig:breakup}
\end{figure}

To illustrate the practical application and value of the traceability framework, the scenario of a new object being discovered following a satellite breakup event is considered. This scenario exemplifies how the system traces back from the identified new object to the initial events that led to its discovery. The sequence of algorithms and data flows involved in this object discovery process is shown in Figure~\ref{fig:data_flow} (Appendix~\ref{sec:data_flow}), and the corresponding decision traceability chain is presented in Figure~\ref{fig:breakup}.

The traceability framework, inspired by MITRE's NIST IR 8536 framework, is implemented by using a message bus, that ensures that every interaction between algorithms is captured as a message. This is an important component that facilitates backwards traces.
\begin{itemize}
    \item \textbf{Message Schema}: Each message contains metadata about the data, including the unique ID of the message, the parent message ID(s), and algorithm that produced the data. This metadata is the key in connecting the messages to each other for traceability.
    \item \textbf{Parent Relationships}: Messages contain links to their parent messages and external data sources, allowing the system to retrace the entire chain of events backwards to the  source.
\end{itemize}

Figure~\ref{fig:breakup} shows the result of this trace back process, where a newly discovered object can be traced back through a number of messages and sub-system components (represented as orange, yellow, and blue circles) to the original task request that triggered the breakup investigation. The green circles represent the messages, which are the results of various processing steps performed by the system. The external data sources consumed by the processing steps are represented as gray circles.
By showing the entire process, in reverse order, the framework provides complete visibility of the data and processes that went into the decision.
Decision traceability for this object discovery scenario provides the following benefits:
\begin{itemize}
    \item \textbf{Auditing}: Enables system engineers to audit the decision paths backwards, verifying that the system made the correct decision by reviewing the input data and the various models that contributed to the final decision.
    \item \textbf{Error Diagnosis}: If a questionable or incorrect object is discovered, operators can use the chain to quickly trace back to the point at which the problem occurred; for instance, a model may have generated a questionable output, or the wrong data was used.
    \item \textbf{Continuous Improvement}: The insight gained from the traceable data and processes allows for continuous improvement by refining the models and the data, and by understanding the impact on the decision, which will ultimately result in more reliable outcomes in the future.
\end{itemize}

\section{Next Steps}

To implement data and decision traceability, the following steps are being pursued:

\begin{itemize}
    \item \textbf{Integration with the message bus}: A key objective of this work is to establish comprehensive traceability within the BMS system. The immediate next step is the onboarding of all algorithms the message bus, which will enable the automatic capture of data related to their inputs and outputs. Given our existing system integration with the message bus, this data capture will begin as soon as each algorithm is integrated. This data will not only enable post-hoc analysis of system behavior but also provide valuable insights for future system optimization and development.
    \item \textbf{Graph Database (DB)}:  To facilitate comprehensive traceability and auditing capabilities within the BMS system, metadata derived from the message bus will be stored in a graph database (DB). Graph databases are particularly well-suited for this purpose due to their ability to efficiently represent and manage complex relationships. By storing metadata as a graph, we can effectively reconstruct the flow of information and the sequence of processing steps that lead to specific outcomes. 
    \item \textbf{Traceability User Interface (UI)}: To facilitate user access to the traceability data, a User Interface (UI) is being developed (Fig.~\ref{fig:UI)} that will allow users to select a specific decision or event and visualize its corresponding traceability chain. The UI employs graphical representations, such as directed graphs or flow diagrams, to effectively convey the complex relationships between data elements and processes. This visualization enables users to quickly grasp the sequence of events and identify critical dependencies, facilitating efficient analysis and understanding of system behavior.
\end{itemize}

\begin{figure}[h]
    \centering
    \includegraphics[width=0.6\linewidth]{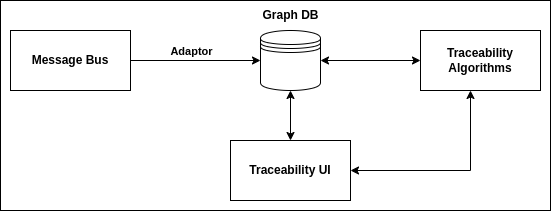}
    \caption{System architecture of the Traceability framework}
    \label{fig:arch}
    \vspace{-5mm}
\end{figure}

\begin{figure}
    \centering
    \includegraphics[width=0.9\linewidth]{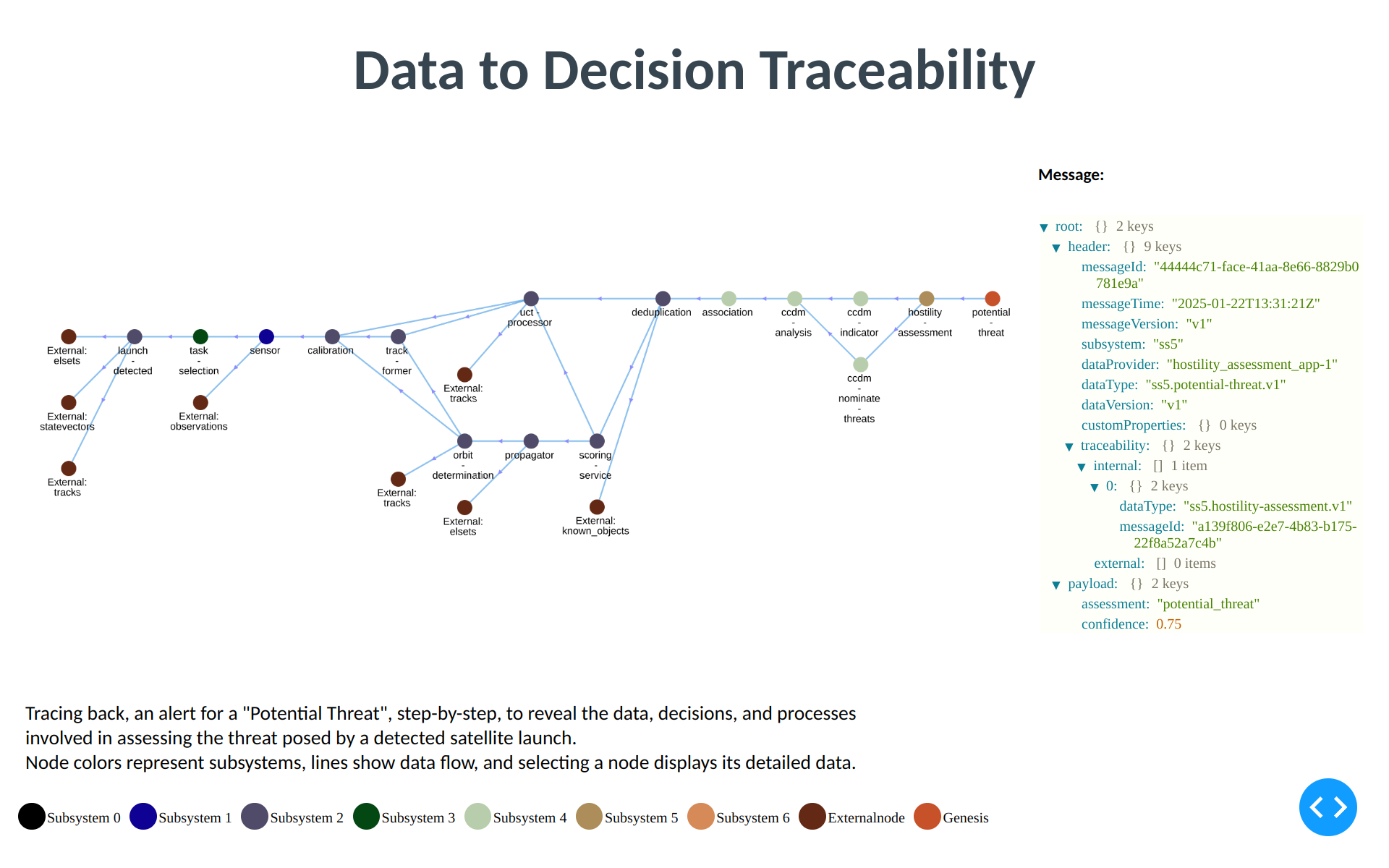}
    \caption{Proposed User Interface (UI) for data to decision traceability capable of demonstrating the algorithms and messages ingested.}
    \label{fig:UI}
\end{figure}

\bibliography{root}
\bibliographystyle{abbrv}

\newpage
\appendix
\section{Reference Algorithms and Data Flow}\label{sec:data_flow}

\begin{figure}
    \centering
    \includegraphics[width=0.6\linewidth]{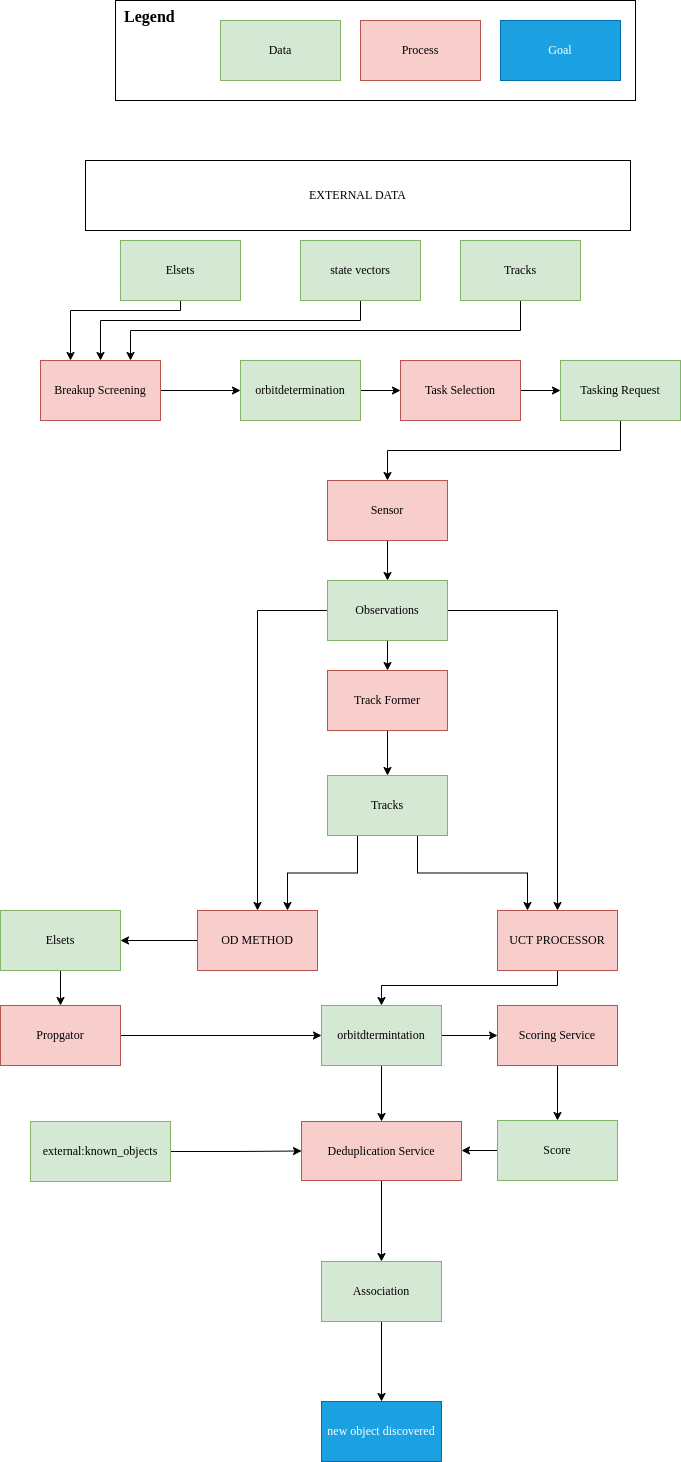}
    \caption{A small scale data and algorithm flow that approximates the message bus.}
    \label{fig:data_flow}
\end{figure}

Figure~\ref{fig:data_flow} presents a simplified data and algorithm flow that illustrates how the message bus functions as the central communication mechanism within the Welder's Arc system. In this diagram, ‘Data’ boxes represent specific data elements, such as tracks and observations. ‘Process’ boxes represent computations or algorithms, like orbit determination and breakup screening. And ‘Goal’ boxes represent a specific outcome in the process, such as discovering a new object. All interactions between these elements – data transfer, algorithm invocation, and state changes – occur through the message bus, enabling the system's data traceability.
\newpage
\section{Message Schema}
Listing~\ref{lst:header} presents an example of a message header incorporating the traceability section. To facilitate comprehensive tracking of data provenance, parent relationships are categorized as either 
\texttt{internal} or \texttt{external}. Internal parent relationships denote data originating within the BMS system via the message bus, while external relationships represent data sources outside WA's boundaries, such as API calls and bulk data requests. For each identified parent, the message header records the data type and the unique message ID. Additionally, a designated \texttt{parameters} field is included to accommodate supplementary information for external resources, such as API call parameters or other relevant metadata.

\begin{lstlisting}[
    language=json, 
    caption=Example Header for Calibration Message,
    label=lst:header
    ]
{
  "messageId": "a47c5d8f-3e2b-4f5a-9825-7ac8f11e2345",
  "messageTime": "2024-12-06T12:34:56Z",
  "messageVersion": "0.0.1",
  "subsystem": "ss0",
  "dataProvider": "spaceprotocol",
  "dataType": "ss0.data.eoobservation.calibrated",
  "dataVersion": "0.1.0",
  "dataPayload": {
    "satnum": "10121",
    "observation": {"x": 0.1234, "y": -0.3456, "t": 1733428563},
    "reference": [
      {"x": 0.1235, "y": -0.3457, "t": 1733427554},
      {"x": 0.1236, "y": -0.3458, "t": 1733427555},
      {"x": 0.1237, "y": -0.3459, "t": 1733427556},
      {"x": 0.1238, "y": -0.3460, "t": 1733427557},
      {"x": 0.1239, "y": -0.3461, "t": 1733427558}
    ],
    "calibrated": {"x": 0.1236, "y": -0.3456x, "t": 1733428563},
  },
  "messageHash" : "35443c...d400",
  "traceability": {
    "internal": [
      {
        "dataType": "ss0.data.eoobservation",
        "messageId": "44444c71-face-41aa-8e66-8829b0781e9a",
        "messageHash": "c079753f...3b9e",
      }
    ],
    "external": [
      {
        "resourceLink": "https://api.spaceprotocol.org/sensor_calibration",
        "parameters": {
          "queryParameter": {
              "sensorid":"Panopticon_012",
              "timestamp":"2024-12-06T12:34:56Z"
          },
        },
        "resourceHash": "4fd5920...2bea",
      }
    ]
  },
}
\end{lstlisting}

\end{document}